\documentclass[prl,a4paper,aps,twocolumn,superscriptaddress,groupedaddress]{revtex4}
\usepackage[dvips]{graphicx}
\usepackage{ae}
\usepackage[T1]{fontenc}
\usepackage[ansinew]{inputenc}
\usepackage{amsmath}
\usepackage{amssymb}
\usepackage{graphicx}
\usepackage{caption}
\usepackage{color}
\usepackage[colorlinks]{hyperref}
\usepackage{lscape}
\usepackage{amsfonts}
\usepackage{amssymb}
\usepackage{makeidx}
\usepackage{braket}
\usepackage{bm}
\usepackage{bbm}
\usepackage{hyperref}
\usepackage{dsfont}
\usepackage{enumerate} 
\usepackage{float} 
\usepackage{mathrsfs}
\usepackage{mathtools}
\usepackage{xfrac}
\usepackage{soul}

\begin{document}
\title{Violation of equivalence in an accelerating atom-mirror system in the generalized uncertainty principle framework }

\author{Riddhi Chatterjee}
\email{riddhi.chatterjee@bose.res.in}

\author{Sunandan Gangopadhyay}
\email{sunandan.gangopadhyay@gmail.com}

\author{A. S. Majumdar}
\email{archan@bose.res.in}
\affiliation{S.N. Bose National Centre for Basic Sciences, Block JD, Sector III, Salt Lake, Kolkata 700106, India.}

\begin{abstract}
We study the spontaneous excitation of a two-level atom in the presence of a perfectly reflecting mirror, when the atom, or the mirror, is uniformly accelerating in the framework of the generalised uncertainty principle (GUP). The quantized scalar field  obeys a modified dispersion relation
leading to a GUP deformed Klein-Gordon equation. The solutions of this equation with suitable boundary conditions are obtained to calculate the spontaneous excitation probability of the atom for the two separate cases. We show that in the case when the mirror is accelerating, the GUP modulates the spatial oscillation of the excitation probability of the atom, thus breaking the symmetry between the excitation of an atom accelerating relative to a stationary mirror, and a stationary atom excited by an  accelerating mirror. An explicit violation of the equivalence principle
seems to be thus manifested. We further obtain an upper bound on the GUP parameter using standard values of the system parameters.
\end{abstract}



\maketitle

\noindent {\it Introduction} : The general theory of relativity discovered by Einstein \cite{einst} is a theory of gravity that has its foundations based on geometrical ideas. It has been realized that there is a deep connection of general relativity with thermodynamics. This understanding has arisen due to 
developments such as the formulation of black hole thermodynamics \cite{therm}, emission of all species of particles from the strong gravitational field of black holes known as Hawking radiation \cite{hawk}, the Unruh effect where accelerating atoms in their ground states moving through Minkowski vacuum go to an excited state by absorbing Rindler particles \cite{unruh}, and an acceleration radiation where an inertial observer interprets the absorption of Rindler particles as the emission of Minkowski particles \cite{accrd,unruhdet}. 

It is also known that the unification of quantum mechanics with the general theory of relativity poses formidable challenges, and the search of a consistent quantum theory of gravity has been one of the main lines of research in theoretical physics. A lot of effort has been devoted in the domains of string theory \cite{string} and loop quantum gravity \cite{loop} to develop a consistent quantum theory of gravity. A common understanding from these investigations has been the existence of an observer independent length scale, the so called Planck length $l_P = \sqrt{G\hbar/c^3}\sim 10^{-33}$ cm. The presence of such a length scale emerges naturally from a modified version of the uncertainty principle known as the generalised uncertainty principle (GUP) \cite{gup}.

The GUP has been employed to address several physical problems, namely, violation of Lorentz invariance \cite{lorentz}, black hole physics \cite{bh}, Unruh effect \cite{undsp}, as well as certain phenomena in low energy systems \cite{low}. Such low energy systems have opened up a new arena to look for indirect experimental evidence of quantum gravity effects \cite{gupexp}. One such evidence is the violation of the gravitational weak equivalence principle. Violation  of the classical weak equivalence principle \cite{cwep} has been extensively studied in arenas such as gravity induced interference experiments \cite{cow}, for particles bound in an external gravitational potential \cite{bound}, and for particles freely falling under gravity \cite{free}, to name a few. The transformation of quantum states between reference frames needs
careful attention \cite{bsu}, leading to implications on the equivalence of
acceleration and gravity for quantum systems \cite{multi}. A quantum version of  the equivalence principle has been proposed in \cite{bruk}.

The  equivalence principle has turned out to be a subject of fascinating debate in the context of terrestrially implementable low energy experiments such as the case of an accelerated atom interacting with a quantum field \cite{sbfp}. 
Investigations have shown that a violation of the principle of equivalence
can be observed in the response function of an Unruh-DeWitt detector for different spacetimes and vacua \cite{dougl}, though the detector's design
and the properties of the field with which it is interacting may also lead to
the absence of thermal response \cite{crispino}.
Radiative properties of single \cite{single} and entangled \cite{entangled} accelerated atoms have been studied earlier extensively. Recently, it has been observed \cite{sbfp} that virtual transitions such as the emission of a real photon from the excitation of a static two-level atom due to the uniform acceleration of a mirror  have a probability governed by the Planck factor which involves the photon frequency and the Unruh temperature. It was further noticed that the result differs from the Unruh radiation of an atom accelerating uniformly with respect to a static mirror. In the latter case, the Planck factor appearing in the probability of transition depends on the frequency of the atom  instead of the frequency of the photon emitted. \par

A subtle manifestation of the equivalence principle occurs nonetheless, in terms of the symmetry between the excitation of a stationary atom by a uniformly accelerating mirror in Minkowski 
spacetime and an atom accelerating uniformly with respect to a stationary mirror  when the frequency of the emitted photon is identical to the frequency of the atom.   The equivalence principle may be envisaged \cite{sbfp,ful} in terms of a symmetry between excitation of a stationary atom by an accelerating mirror (Rindler vacuum) and the excitation of an atom freely falling under gravity with respect to a stationary mirror (Boulware vacuum). A similar argument can be put forward in terms of symmetry between the excitation of an atom accelerating in Minkowski spacetime relative to a stationary mirror (Minkowski vacuum) and the excitation of a stationary atom by a mirror freely falling in a gravitational field (Hartle-Hawking vacuum). A deviation from such symmetry can be regarded as a manifestation of violation of the equivalence principle.

The relevance of the GUP comes precisely in this context, since it takes 
quantum gravity effects into account. However, violation of the equivalence principle in the framework
of the GUP has hitherto remained unexplored in the literature. In the present work our main goal is to look at the status of the equivalence principle in the 
framework of the GUP. Our investigation is performed in the setting of a
relatively accelerating atom-mirror system. Here, it is worthwhile to note that 
experimental implementation of the accelerating atom-mirror system has
been proposed using superconducting circuits \cite{wil1,su,joh,luk}.
Moreover, efforts to  constrain the value of the GUP parameter that have so far come from nano-mechanical and opto-mechanical set-ups \cite{bawaj}, can
easily be extended for the present set-up, as well. 

{\it Excitation of an atom by different vacua in the GUP framework} :
The simplest form of the GUP proposed in the literature reads \cite{low}
\begin{eqnarray}
\Delta q_{i} \Delta p_{i}\geq \frac{\hbar}{2}\left[1+\beta(\Delta p ^2 +
\langle p\rangle ^2)+2\beta(\Delta p_{i}^2 +
\langle p_i\rangle ^2)\right]
\label{unc1}
\end{eqnarray}
where $\beta$ is the GUP parameter ($\beta \to 0$ leads to the limit of the
Heisenberg uncertainty relation), and  $p^2 =p_{i}p_{i}$ (with sum on $i$) and $i=1,2,3$. This is equivalent
to the following modified Heisenberg algebra 
\begin{eqnarray}
[q_i, p_j]=i\hbar(\delta_{ij} +\beta \delta_{ij}p^2 +2\beta p_i p_j).
\label{unc2}
\end{eqnarray}
 We consider an atom-mirror system in the presence of a quantized scalar field whose canonical momentum operator satisfies the GUP modified dispersion relation. The atom is assumed to have two energy levels $\lbrace \ket{g}, \ket{e} \rbrace$ and energy eigenvalues $\lbrace - \frac{\omega_0}{2}, \frac{\omega_0}{2} \rbrace$. We  study the spontaneous excitation of the atom along with the simultaneous emission of a photon when the atom and the mirror are in relative acceleration, in framework of the GUP.  Atomic excitation occurs due to the excitation of the quantum field vacuum due to acceleration. Our aim is to find the effects of the GUP on the transition probabilities of the virtual transitions that can occur, and then look at the status of the equivalence principle. We consider here the
set-up of a single mode cavity to obtain explicitly the spatial dependence of the interference pattern exhibited by the transition probailities \cite{sbfp}.

{\it Atom accelerating away from static mirror} :
First, we consider the situation when the mirror is static at a spatial position $z = z_0$ in the Minkowski spacetime $(t,z)$. The modified Klein-Gordon equation in $(1+1)$-dimensions reads~\cite{undsp}
\begin{align} \label{klein}
\Big(\frac{1}{c^2} \  \partial_t^2 - \partial_z^2 + 2 \beta \hbar^2 \  \partial_z^4 \Big) \phi (t,z) = 0
\end{align}
The field $\phi$ satisfies the boundary condition $\phi (t, z_0) = 0$. We now take the solution of the above equation in the form $\phi_{\nu}(t,z) = e^{-i\nu t}  e^{nz}$, where $\nu (>0)$ is the frequency of the photon. Substituting this in the above equation, we get \begin{align}
n^2 - 2 \beta \hbar^2 n^4 + \frac{\nu^2}{c^2} = 0 .
\end{align}
To solve the above equation, we take $n = (n_0 + \beta \  \widetilde{n})$, where $n_0$
and $\widetilde{n}$ are to be determined. Substituting this in the above equation and comparing coefficients of powers of $\beta$ upto $\mathcal{O}(\beta)$ on both sides of the equation, we find $n_{0}=\pm i\nu/c$ and $\widetilde{n}=\mp i\hbar^2 \nu^3/c^3$. This then gives
\begin{align}
n = \pm\ i \frac{\nu}{c}\Big(1- \beta \hbar^2 \frac{\nu^2}{c^2}\Big).
\end{align}
Hence, the solution of eq.(\ref{klein}) that satisfies the boundary condition $\phi (t,z_0) = 0$ is given by 
\begin{align}\label{field1}
\phi_{\nu} (t,z) = e^{-i\nu t}  e^{- i\frac{\nu}{c}(1-\beta \hbar^2 \frac{\nu^2}{c^2})(z-z_0)} \nonumber \\ -\ e^{-i\nu t}  e^{i\frac{\nu}{c}(1-\beta \hbar^2 \frac{\nu^2}{c^2})(z-z_0)}.
\end{align}
We take  the atom to accelerate along the positive $z$-direction with acceleration $a$.  The trajectory of the atom is of the form
\begin{align}\label{rind1}
t(\tau) = \frac{c}{a} \sinh{\Big(\frac{a\tau}{c}\Big)} \qquad z(\tau) = \frac{c^2}{a} \cosh{\Big(\frac{a\tau}{c}\Big)}
\end{align}
where $\tau$ is the proper time of the atom.

The atom-field interaction Hamiltonian is given by \begin{align}\label{ham}
H_I(\tau) = \hbar g (\hat{a}^{\dagger}_{\nu} \phi^{\ast}_{\nu} (t,z) + \hat{a}_{\nu} \phi_{\nu} (t,z)) \cdot \nonumber \\ \frac{i}{2}(\ket{g}\bra{e} e^{-i\omega_0 \tau} - \ket{e}\bra{g} e^{i\omega_0 \tau})
\end{align}
where $g$ is the atom-field coupling constant, which is assumed to be independent of $\tau$. $\hat{a}_{\nu},\hat{a}^{\dagger}_{\nu}$ are the annihilation and creation operators of the scalar field. The atomic transition amplitude is given by \begin{align}\label{amp}
\mathcal{A} = \frac{1}{\hbar}\int d\tau \braket{1_{\nu},e \mid H_I(\tau) \mid 0,g}.
\end{align}
Hence, the atomic transition probability is given by 
\begin{eqnarray}
P_1 &=&  \frac{1}{\hbar^2} \Bigg\lvert \int d\tau \braket{1_{\nu},e \mid H_I(\tau) \mid 0,g}\Bigg\rvert ^2 \nonumber \\
&=& \frac{g^2}{4} \Bigg\lvert \int_{-\infty}^{\infty} d\tau \ \phi_{\nu}^{\ast}(t,z) e^{i\omega_0 \tau} \Bigg\rvert ^2.
\end{eqnarray}
We now plug in the expression of $\phi_{\nu}^{\ast}(t,z)$ from eq.(\ref{field1}) and substitute the atomic trajectories from eq.(\ref{rind1}) in the above equation. After some calculation, we get 
\begin{eqnarray}
P_1 &=& \frac{g^2}{4} \Bigg\lvert \int_{-\infty}^{\infty} d\tau \Big[ e^{( \frac{i\nu}{c}\alpha_1 e^{\frac{a\tau}{c}} -i\alpha_2 e^{- \frac{a\tau}{c}} -\frac{i\nu}{c}\alpha_3 z_0)} \nonumber \\ &-& e^{-(\frac{i\nu}{c}\alpha_1 e^{-\frac{a\tau}{c}} -i\alpha_2 e^{\frac{a\tau}{c}} -\frac{i\nu}{c}\alpha_3 z_0)} \Big] e^{i\omega_0 \tau} \Bigg\rvert ^2
\end{eqnarray}
where $\alpha_1 = (1- \frac{\beta \hbar^2 \nu^2}{2 c^2})$, $\alpha_2 = \frac{\beta \hbar^2 \nu^3}{2 a c}$, $\alpha_3 = (1- \frac{\beta \hbar^2 \nu^2}{c^2})$. Evaluating the above integral, we get 
\begin{eqnarray}\label{prob1}
&P_1& = \frac{2\pi g^2 c}{a\omega_0}\cdot \frac{e^{-\big(\frac{\beta \hbar^2 \nu^4}{a^2}\Omega \cos{\Delta}\big)}}{e^{(\frac{2\pi \omega_0 c}{a})} - 1} \nonumber \\ & \times & \sin^{2}{\Big(\frac{\tilde{\nu}z_0}{c} - \eta - \frac{\beta \hbar^2 \nu^2 \omega_0}{2ac} + \dfrac{\beta \hbar^2 \nu^4}{2a^2}\Omega \sin{\Delta}\Big)}\quad
\end{eqnarray}
where $\tilde{\nu} = \nu \Big(1- \frac{\beta \hbar^2 \nu^2}{c^2}\Big)$,
$\eta = \delta + \theta$, $\delta = \frac{\omega_0 c}{a} \ln{\frac{a}{\nu c}}$,
$\theta = Arg\Big(\Gamma \Big(-\frac{i\omega_0 c}{a}\Big) \Big)$,
$\theta_1 = Arg\Big(\Gamma \Big(-\frac{i\omega_0 c}{a} - 1\Big) \Big)$,
$\Delta = (\theta_1 - \theta)$, and 
$\Omega = \frac{\big\lvert \Gamma \big(-\frac{i\omega_0 c}{a} - 1\big)\big\rvert}{\big\lvert \Gamma \big(-\frac{i\omega_0 c}{a}\big)\big\rvert}$~. From the above expression, we observe that the Planck factor is governed by the frequency of the atom. However, the interference term arising due to the incident and reflected waves contain the effect of the GUP.

{\it Mirror accelerating away from static atom} :
We next consider the case where the atom is static at spatial position $z = z_0 < c^2/a$ in the Minkowski spacetime. 
The mirror accelerates away from the atom with an acceleration $a$. The presence of the accelerating mirror modulates the field mode. The trajectory of the accelerated mirror is given by eq.(\ref{rind1}). Coordinate transformation between the frame of mirror, that is, the Rindler frame $(\bar{t},\bar{z})$ and the Minkowski frame reads
\begin{align}\label{rind2}
t = \frac{c}{a} e^{\sfrac{a\bar{z}}{c^2}} \sinh{\Big(\frac{a\bar{t}}{c}\Big)} \qquad z = \frac{c^2}{a} e^{\sfrac{a\bar{z}}{c^2}} \cosh{\Big(\frac{a\bar{t}}{c}\Big)}~.
\end{align}
The mirror is spatially static in the Rindler frame and its trajectory is given by $\bar{z} = 0$. 
The quantized scalar field in Rindler spacetime satisfies the modified Klein-Gordon equation given by eq.(\ref{klein}) with $(t,z)$ replaced by $(\bar{t},\bar{z})$ and boundary condition $\phi (\bar{z}) = 0$. Its solution is given by \begin{align}\label{field2}
\phi_{\nu} (\bar{t},\bar{z}) = e^{-i\nu \bar{t}}  \left(e^{i\frac{\nu}{c}(1-\beta \hbar^2 \frac{\nu^2}{c^2})\bar{z}}  -  e^{- i\frac{\nu}{c}(1-\beta \hbar^2 \frac{\nu^2}{c^2})\bar{z}}\right).
\end{align}
We now need the inverse transformations of eq.(\ref{rind2}) which are given by \begin{eqnarray}
\bar{t}(t,z) &=& \frac{c}{2a}\ln{\left( \frac{z+ct}{z-ct}\right)}\nonumber \\
\bar{z}(t,z) &=& \frac{c^2}{2a}\ln{\left(\frac{a^2}{c^4}\left(z^2 - c^2t^2\right)\right)}.
\end{eqnarray}
The above transformations are defined for $z>c\lvert t\rvert$. 
Substituting these transformations in eq.(\ref{field2}) and simplifiying, we get the field mode in the Minkowski spacetime as
\begin{align}\label{field3}
&\phi (t,z) = e^{i\left[\frac{\bar{\nu}c}{a}\ln{\left(\frac{a}{c^2}(z-ct)\right)}\right]} \left(\frac{a}{c^2}\left(z+ct\right)\right)^{-\frac{i\beta \hbar^2\nu^3}{2ac}} \Theta(z-ct) \nonumber \\ &- e^{-i\left[\frac{\bar{\nu}c}{a}\ln{\left(\frac{a}{c^2}(z+ct)\right)}\right]} \left(\frac{a}{c^2}\left(z-ct\right)\right)^{\frac{i\beta \hbar^2\nu^3}{2ac}} \Theta(z+ct)
\end{align}
where $\bar{\nu} = (1- \frac{\beta \hbar^2 \nu^2}{2 c^2})\nu$. 
The atom-field interaction Hamiltonian and the transition probability amplitude are  the same as eq.(s)(\ref{ham},\ref{amp}) with $\tau$ replaced by $t$ and $(t,z)$ replaced by $(\bar{t},\bar{z})$. The atomic transition probability evaluated at atomic position $(t,z_0)$ is given by \begin{align}
P_2 = \frac{g^2}{4} \Bigg\lvert \int_{-\infty}^{\infty} dt \ \phi_{\nu}^{\ast}(z_0,t) e^{i\omega_0 t} \Bigg\rvert ^2.
\end{align}
Substituting the expression of $\phi^{\ast}(t,z)$ from eq.(\ref{field3}) in above equation and after some calculation, we get \begin{align}
P_2 = \frac{g^2}{4} \Bigg\lvert \int_{-z_0/c}^{\infty} dt \ e^{-i\left[\frac{\bar{\nu}c}{a}\ln{\left(\frac{a}{c^2}(z_0+ct)\right)} + \omega_0 t\right]}\nonumber \\ \left(\frac{az_0}{c^2}\left(1-\frac{ct}{z_0}\right)\right)^{\frac{i\beta \hbar^2\nu^3}{2ac}} + cc. \Bigg\rvert ^2 ~.
\end{align}
Evaluating the above integral, we get 
\begin{align}\label{prob2}
P_2 &= \frac{2\pi g^2 \bar{\nu}c}{a\omega_0^2}\cdot \frac{e^{-\big(\frac{\beta \hbar^2 \nu^3}{az_0\omega_0}\big)}}{e^{(\frac{2\pi \bar{\nu} c}{a})} - 1} \cdot  \sin^{2}\Big(\frac{\omega_0 z_0}{c} - \frac{\bar{\nu}c}{a}\ln{\left(\frac{a}{\omega_0 c}\right)} \nonumber \\ & + \frac{\beta \hbar^2 \nu^3}{2ac} \ln{\left(\frac{az_0}{c^2}\right)} + \dfrac{\beta \hbar^2 \nu^3}{2ac} + \dfrac{\beta \hbar^2 \nu^4 c}{2a^2 z_0 \omega_0} + \kappa\Big)
\end{align}
where $\kappa = Arg\Big(\Gamma \Big(\frac{i\bar{\nu} c}{a}\Big) \Big) $. 
The Planck factor here is governed by a GUP modified photon frequency.
However, as different from the reference \cite{sbfp}, the spatial oscillation in the interference pattern here is very interesting. Apart from the usual 
position dependent atomic frequency term, there is another position dependent term which depends on the field frequency. This term owes its origin to the GUP. It implies that the interference pattern gets modified by the field frequency in the presence of the GUP. 

{\it Violation of the equivalence principle} :
Let us now set $\omega_0 = \nu$, making the frequencies of the atom and
the photon same. It can be observed that even for this case, 
the spatial oscillations for the two probabilities are not the same, as is evident 
from eq.(s)(\ref{prob1},\ref{prob2}), in contrast to the framework based
on the Heisenberg uncertainty relation \cite{sbfp}. This therefore breaks the symmetry between the excitation of an atom accelerating in Minkowski spacetime relative to a stationary mirror and a stationary atom excited by an uniformly accelerating mirror. This feature can hence be regarded as a manifestation of violation of the equivalence principle originating from the GUP.
It can be checked that by setting $\beta = 0$ in eq.(s)(\ref{prob1},\ref{prob2}), the symmetry ensues, restoring the equivalence in the
Heisenberg uncertainty framework.

There have been proposals to provided bounds 
on the value of the GUP parameter $\beta$ resulting from various effects such as, correction in Lamb shift, Landau Levels, simple harmonic oscillators, and gravitational wave detections \cite{gupexp}. 
Here we provide an estimate of the upper bound on $\beta$ from the exponent of the damping factor. It is clear from the exponential factor in eq.(\ref{prob2}) that in order to ensure that the GUP corrections do not dominate over the results obtained in the Heisenberg uncertainty principle framework, we must have 
$\left(\frac{\beta \hbar^2 \nu^3}{az_0\omega_0}\right) << 1$.
Taking $\nu =\omega_0=1$GHz \cite{sbfp} and $az_0 \sim c^2$, 
 we find  $\beta <<  10^{67}/(M_P c)^2 $, with $M_P$ being the Planck mass. Though this bound is weaker than the bound obtained on $\beta$ in the
context of gravitational waves \cite{sksg}, our result provides an example of the
possibility of formulating testable bounds on the GUP parameter in the
context of controllable low energy atom-photon interactions.
Interestingly, bounds on the GUP parameter also arise from the mismatch in the spatial oscillation of the two probabilities. The ratio between the spatial part of the second and first probabilities is given by
$R = 1 + \mathcal{Q}(z_0)$,
where $\mathcal{Q}(z_0)=\frac{\beta \hbar^2 \nu^2 c^2}{2a^2 z_0^2} + \frac{\beta \hbar^2 \nu^2}{2az_0} \ln{\left(\frac{az_0}{c^2}\right)}$ can be regarded
as an equivalence violation parameter. This provides a similar bound on $\beta$ as obtained above.

{\it Conclusions} :
We now summarize our findings with some observations. The main focus of this paper is to look at the status of the symmetry between the excitation of a stationary atom by an accelerating mirror and a uniformly accelerating atom relative to a stationary mirror taking into account Planck scale effects. Such a symmetry has been shown to be valid in the framework  of the  Heisenberg uncertainty relation, and has been interpreted as a manifestation of the principle of equivalence  in \cite{sbfp,ful}. It should be noted however, that the interpretation of this symmetry as a manifestation of the equivalence principle goes beyond the well known classical version which states that  by local measurements it would be impossible to distinguish between an inertial observer in Minkowski spacetime and a free-falling observer in a gravitational field (or, equivalently, a static observer in a uniform gravitational field and a uniformly accelerated observer in flat spacetime).

Our methodology is to consider a quantized scalar field vacuum that obeys the GUP modified dispersion relation. From the GUP modified Klein-Gordon equation,
we obtain the solutions of the scalar field with particular boundary conditions imposed by the mirror in the two separate cases. Using these solutions we calculate the excitation probabilities of the atom in both the cases, which are found to display significant physical differences. 
 
In the first case, the GUP contributes as a constant phase in the interference. However, in the second case the spatial oscillation gets modified by an additional term containing the field frequency and the GUP parameter $\beta$. Hence, we find that 
the symmetry observed in \cite{sbfp} gets broken in the framework of the GUP
even when $\nu = \omega_0$. This is the most striking result of our study,
and may be interpreted as an explicit violation of the equivalence principle. This is because the symmetry between the excitation of an atom accelerating in Minkowski spacetime relative to a stationary mirror and a stationary atom excited by an uniformly accelerating mirror (considered to be a manifestation of the equivalence principle in \cite{sbfp,ful}) gets broken. Further, using the condition that the GUP
induced corrections do not dominate over the corresponding expressions obtained
using the Heisenberg uncertainty relation, it is possible to constrain the
value of the GUP parameter in the context of this low energy interacting
atom-mirror set-up.
 
Before concluding, it may be noted that in both the cases the excitation probabilities contain a GUP induced damping factor. In the first case the probability is proportional to the Planck factor containing the atomic transition frequency and the Unruh temperature given by $T_U = \frac{\hbar a}{2\pi k_B c}$, that one gets when $\beta =0$. In the second case, the atomic excitation probability is proportional to the Planck factor which is a function of $\bar{\nu} = (1- \frac{\beta \hbar^2 \nu^2}{2 c^2})\nu$.  Thus, in the presence of the GUP modification, the excitation probability is proportional to the Planck factor containing the field frequency and the modified Unruh temperature given by $T'_U = T_U/(1- \frac{\beta \hbar^2 \nu^2}{2 c^2})$. Since the Planck distribution in eq.(\ref{prob2}) is analogous to the photon spectrum of an atom falling freely in the gravitational field of a Schwarzschild black hole \cite{schw}, this implies that the acceleration radiation observed by
  a distant observer will be a thermal distribution with a GUP modified Hawking temperature.




\begin{thebibliography}{100}
\bibitem{einst} A. Einstein, Annalen der Physik, {\bf 49}, 769 (1916).
\bibitem{therm} S. Carlip, Int. J. Mod. Phys. D \textbf{23}, 1430023 (2014).
\bibitem{hawk} S. W. Hawking, Commun. Math. Phys. \textbf{43}, 199 (1975).
\bibitem{unruh} W. G. Unruh, Phys. Rev. D \textbf{14}, 870 (1976).
\bibitem{accrd} W. G. Unruh and R. M. Wald, Phys. Rev. D \textbf{29}, 1047 (1984).
\bibitem{unruhdet} M. S. Soares, N. F. Svaiter, C. A. D. Zarro, and G. Menezes
Phys. Rev. A {\bf 103}, 042225 (2021); V. Sudhir, N. Stritzelberger, and A. Kempf
Phys. Rev. D {\bf 103}, 105023 (2021).

\bibitem{string} K. Becker, M. Becker, J. Schwarz, {\it String theory and M-theory: A modern introduction}, (Cambridge University Press) ISBN 978-0-521-86069-7.
\bibitem{loop} Dah-Wei Chiou, Int. J. Mod. Phys. D \textbf{24}, 1530005 (2015).
\bibitem{gup} D. Amati, M. Ciafaloni, and G. Veneziano, Phys. Lett. B \textbf{216}, 41 (1989); M. Maggiore, Phys. Lett. B \textbf{304}, 65 (1993); M. Maggiore, Phys. Rev. D \textbf{49}, 5182 (1994); M. Maggiore, Phys. Lett. B \textbf{319}, 83 (1993).

\bibitem{lorentz} T. Jacobson and D. Mattingly, Phys. Rev. D \textbf{63}, 041502 (2001); {\it ibid}. \textbf{64}, 024028 (2001).
\bibitem{bh} R. J. Adler, P. Chen and D. I. Santiago , Gen. Rel. Grav. \textbf{33}, 2101 (2001); S. Gangopadhyay, A. Dutta, A. Saha, Gen. Rel. Grav. \textbf{46}, 1661 (2014); A. Dutta, S. Gangopadhyay, Gen. Rel. Grav. \textbf{46}, 1747 (2014); 
S. Gangopadhyay, A. Dutta, M. Faizal, Euro.Phys.Lett. \textbf{112} (2015) 2, 20006;
Z.W. Feng, H.L. Li, X.T. Zu, S.Z.Yang, Eur.Phys. J. C \textbf{76} (2016) 212;
Z.W. Feng, S.Z.~Yang, H.L.~Li,, X.T. Zu, Phys. Lett. B \textbf{768} (2017) 81;
J. Gin\'{e}; Commun. Theor. Phys. \textbf{73} 015201 (2020).



\bibitem{undsp} A. Mukherjee, S. Gangopadhyay, M. Dutta, Euro.Phys.Lett \textbf{129}, 30002 (2020).

\bibitem{low} A. Kempf, G. Mangano, and R. B. Mann, Phys. Rev. D \textbf{52}, 1108-1118 (1995); A. F. Ali, S. Das, and E. C. Vagenas, Phys. Lett. B, \textbf{678} 497-499 (2009).

\bibitem{gupexp} A. F. Ali \textit{et al.}, Phys. Rev. D \textbf{84}, 44013 (2011); I. Pikovski \textit{et al.}, Nature Physics \textbf{8}, 393-397 (2012); P. Bosso \textit{et al.}, Phys. Rev. A \textbf{96}, 023849 (2017); 
Z.-W. Feng, S.-Z. Yang, H.-L. Li, X.-T. Zu, Phys. Lett. B \textbf{768}, 81 (2017);
P. Bosso \textit{et al.}, Phys. Lett. B \textbf{785}, 498-505 (2018).
\bibitem{cwep} P. Holland,   {\it The Quantum Theory of Motion} (London: Cambridge University Press) pp 259-66 (1993).
\bibitem{cow} R. Colella, A. W. Overhauser, S. A. Werner,  Phys. Rev. Lett. \textbf{34}, 1472 (1975); A. Peters, K. Y. Chung, S. Chu, Nature \textbf{400}, 849 (1999).
\bibitem{bound} D. M. Greenberger, A. W. Overhauser, Rev. Mod. Phys. \textbf{51}, 43 (1979); D. M. Greenberger, Ann. Phys. \textbf{47}, 116 (1968); Rev. Mod. Phys. \textbf{55}, 875 (Rev. Mod. Phys. 55 875).
\bibitem{free} L. Viola, R. Onofrio, Phys. Rev. D \textbf{55}, 455 (1997); M. A. Ali, A. S. Majumdar, D. Home, A. K. Pan, Class. Quantum Grav. \textbf{23}, 6493 (2006); P. Chowdhury, D. Home, A. S. Majumdar, S. V. Mousavi, M. R. Mozaffari, S. Sinha, Class. Quantum Grav. \textbf{29}, 025010 (2012); S. V. Mousavi, A. S. Majumdar, D. Home, Class. Quantum Grav. \textbf{32}, 215014 (2015).
\bibitem{bsu} J. S. Ben-Benjamin, M. O. Scully, W. G. Unruh, Phys, Scr. {\bf 95}, 074015 (2020).
\bibitem{multi} J. S. Ben-Benjamin, M. O. Scully, S. A. Fulling, {\it et al.}. Int. J. Mod. Phys. A {\bf 34}, 1941005 (2019).
\bibitem{bruk} M. Zych, C. Brukner, Nature Phys \textbf{14}, 1027 (2018). 

\bibitem{sbfp} A. A. Svidzinsky, J. S. Ben-Benjamin, S. A. Fulling, D. N. Page, Phys. Rev. Lett. \textbf{121}, 071301 (2018).


\bibitem{dougl}D.~Singleton, S.~Wilburn, Phys.Rev.Lett. \textbf{107} (2011) 081102;
L.C.B. Crispino, A. Higuchi, G.E.A. Matsas, Phys.Rev.Lett. \textbf{108} (2012) 049001;
D.~Singleton, S.~Wilburn, Phys.Rev.Lett. \textbf{108} (2012) 049002;
Z. Foo, S. Ono, M. Zych, Phys. Rev. D \textbf{102}, 085013 (2020).

\bibitem{crispino} L.C.B. Crispino, A. Higuchi, G.E.A. Matsas, Rev. Mod. Phys.
{\bf 80}, 787 (2008).


\bibitem{single} J. Audretsch and R. M\"{u}ller, Phys. Rev. A \textbf{50}, 1755 (1994); H. Yu and S. Lu, Phys. Rev. D \textbf{72}, 064022 (2005);  Phys. Rev. D \textbf{73}, 109901 (2006).
\bibitem{entangled} L. Rizutto \textit{et al.}, Phys. Rev. A \textbf{94}, 012121 (2016); W. Zhou, R. Passante, L. Rizzuto, Symmetry \textbf{10}, 185 (2018);  W. Zhou, H. Yu, Phys. Rev. D \textbf{101}, 085009 (2020); R. Chatterjee, S. Gangopadhyay, A. S. Majumdar,  Eur.Phys.J.D \textbf{75} (2021) 6, 179.




\bibitem{ful} S. A. Fulling and J. H. Wilson, Physica Scripta \textbf{94} (1), 014004 (2019).

\bibitem{wil1} C. M. Wilson, G. Johansson, A. Pourkabirian, M. Simoen, J. R. Johansson, T. Duty, F. Nori, and P. Delsing, Nature (London) \textbf{479}, 376 (2011); P. Lahteenmaki, G. S. Paraoanu, J. Hassel, and P. J. Hakonen, Proc. Natl. Acad. Sci. U.S.A. \textbf{110}, 4234 (2013).
\bibitem{su} D. Su, C. T. M. Ho, R. B. Mann, and T. C. Ralph,  New J. Phys. \textbf{19}, 063017 (2017).
\bibitem{joh} J. R. Johansson, G. Johansson, C. M. Wilson, and F. Nori, Phys. Rev. Lett. \textbf{103}, 147003 (2009).
\bibitem{luk} M. Wallquist, K. Hammerer, P. Rabl, M. Lukin, and P. Zoller,  Phys. Scr. \textbf{T137}, 014001 (2009).

\bibitem{bawaj} M. Bawaj et. al., Nat. Commun. {\bf 6}, 7503 (2015); P. Girdhar, A. Doherty,
New J. Phys. {\bf 22},  093073 (2020).

\bibitem{sksg}S.~Bhattacharyya, S.~Gangopadhyay, A.~Saha,
Class. Quantum Grav. {\bf{37}} (2020) 195006.
\bibitem{schw} M. O. Scully, S. Fulling, D. Lee, D. Page, W. Schleich, and A. A. Svidzinsky, Proc. Natl. Acad. Sci. U.S.A. \textbf{115} (32),  8131  (2018).

\end{thebibliography}
\end{document}